\documentclass[prx,amssymb,reprint,superscriptaddress,showpacs]{revtex4-1}
\usepackage{amsmath}
\usepackage{amsfonts}
\usepackage{amssymb}
\usepackage{epsfig}
\usepackage{graphicx}

\renewcommand{\phi}{\varphi}

\newcommand{\be}{\begin{equation}}
\newcommand{\ee}{\end{equation}}
\newcommand{\bea}{\begin{equnaray}}
\newcommand{\eea}{\end{equnaray}}
\newcommand{\ba}{\begin{align}}
\newcommand{\ea}{\end{align}}

\usepackage{color}

\begin{document}

\title{Excitation of vibrational soft modes in disordered systems using active oscillation}
\author{Elsen Tjhung}
\affiliation{Laboratoire Charles Coulomb, UMR 5221, CNRS and Universit\'e Montpellier, Montpellier 34095, France.}
\altaffiliation{Present address: Department of Applied Mathematics and Theoretical Physics, Centre for Mathematical Sciences, University of Cambridge, Wilberforce Road, Cambridge CB3 0WA, United Kingdom.}

\author{Takeshi Kawasaki}
\email{Email: kawasaki@r.phys.nagoya-u.ac.jp} 
\affiliation{Department of Physics, Nagoya University, Nagoya 464-8602, Japan.}

\date{\today}

\begin{abstract}
We propose a new method to characterize the spatial distribution of particles' vibrations in solids with much lower 
computational costs compared to the usual normal mode analysis.
We excite the specific vibrational mode in a two dimensional athermal jammed system
by giving a small amplitude of active oscillation to each particle's size with an identical driving frequency.
The response is then obtained as the real time displacements of the particles. 
We show remarkable correlations between the real time displacements and 
the eigen vectors obtained from conventional normal mode analysis.
More importantly, from these real time displacements, 
we can measure the participation ratio and spatial polarization of particles' vibrations.
From these measurements, we find three distinct frequency regimes
which characterize the spatial distribution and correlation of particles' vibrations in jammed amorphous solids.
Furthermore, we can possibly apply this method to a much larger system to examine the low frequency behaviour 
of amorphous solids with a much higher resolution of the frequency space.

\end{abstract}

\pacs{05.10.-a,61.43.-j}


\maketitle
\section{Introduction}
Vibrational properties in amorphous solids are ill-understood compared to those in crystalline solids.
In crystalline solids, 
the Debye model explains that their vibrational density of states (VDOS) is expressed as  
$D(\omega) \propto \omega^{d-1}$ where $\omega$ is vibrational frequency and $d$ is dimensionality~\cite{AshcroftMermin_book}. 
On the other hand, in amorphous solids, 
the low frequency VDOS shows excess mode spectra over that obtained from the Debye model. 
Although such low frequency vibrational modes are responsible for the anomalous dynamics and statics in 
amorphous materials, their physical origins remain elusive.

Such low frequency excess vibrational modes are most often observed in athermal jammed solids~\cite{ohern, wyart05, hecke, lerner}.
The jammed solids are categorized as one class of amorphous solids.
Experimentally, such jammed solids are, in many cases, composed of soft materials such as 
emulsions~\cite{mason}, foams~\cite{hertzhft} and large colloids~\cite{wagner_book, boyer}.
Interestingly, the VDOS in such systems shows a plateau of modes spectra in the low frequency region 
which extends to $\omega=0$ as the volume fraction approaches the jamming transition point from above~\cite{ohern, wyart05,lerner}. 
Furthermore the low frequency vibrational modes in athermal jammed solids
(which are called soft modes) are also highly non-trivial. 
Here, some modes appear to be collective and extended throughout the space and others are quasi-localized~\cite{hecke,manning}, as in low temperature glasses~\cite
{elliot,schober91,schober93,schober96,harrowell08,elliott99,harrowell09,chen10}.
To characterize the spatial distribution and correlation of such vibrational modes is significant in jammed soft materials.
Actually, it has numerous applications such as
predicting where plastic deformations (leading to mechanical failure) might occur in space, which is called soft spot~\cite{manning}.

To obtain the spatial distribution and correlation of vibrational eigen modes
in dense particle assemblies, 
normal mode analysis is frequently used~\cite{AshcroftMermin_book}. 
In such analysis, the vibrational modes are approximated as harmonic oscillators.
Here, they are characterized by diagonalizing the Hessian matrix, 
which is composed of second derivatives of the potential energy with respect to the coordinates of the particles. 
Unfortunately, its computational and memory costs tend to be huge,
because even when rotational motions are neglected, the matrix size is $dN\times dN$, where  $N$ is the particle number.  
Thus in previous studies~\cite{schober91,schober93,schober96,elliott99,harrowell08,harrowell09,ohern,wyart05,hecke,lerner},
small system sizes tend to be used.
On the other hand, in order to study very low frequency modes, larger system sizes will be inevitable, and thus, 
another innovative method with low computational costs will be desired.

To characterize these spatial properties, the participation ratio 
is often measured~\cite{schober91,schober93,schober96,elliott99,chen10,harrowell09}.
On the other hand, the participation ratio alone is not sufficient to characterize the degree of collectivity in the particles' vibrations
-- it only measures the degree of localization in the particles' vibrations.
Thus, in this study, we shall introduce the average local polarization as a measure of collectivity in the vibrations of the particles.
 
In this paper, we outline a novel method to characterize the spatial distribution and correlation of vibrational modes without diagonalizing a huge Hessian matrix. 
Although our case study is a small system size of $N=1000$, this method can be easily extended to much 
larger system sizes.
It should also be noted that our method cannot directly obtain VDOS, it only tells us the spatial properties of the 
vibrational modes.
A method to obtain VDOS without computing Hessian matrix has already been 
described~\cite{rahman76,rahman81,ikedaJCP12,shintani08}.
Therefore, combined with our new method, we can obtain most of the basic vibrational quantities in stead of the 
normal mode analysis.
In our method, we excite a specific vibrational mode by giving active oscillation in each particle's diameter with an 
identical driving frequency.
The response is then obtained as the real time displacements of the particles, 
which can be compared to the conventional normal modes obtained from static Hessian matrix.
We show high degree of correlation between the real time displacements and the static normal modes.
Furthermore, we also characterize the spatial structure of these vibrational normal modes 
by measuring the participation ratio~\cite{schober91,schober93,schober96,elliott99,chen10,harrowell09}
and the polar order parameter.
The results of these measurements are again found to be consistent for both the real time displacements and the static 
normal eigen mode.
More importantly, from these measurements we can distinguish three distinct frequency regimes based on the spatial 
distribution of the particles' vibrations (whether they are extended and/or collective in space).
Finally by directly measuring the real time displacements of the particles, 
we may obtain dynamical quantities such as the mean squared displacement (MSD). 
The MSD gives us the relative amplitude of particles' vibrations. 
In particular we find that particles vibrate over larger distances at lower frequency excitation.
This is consistent to the experiments of colloidal particles~\cite{tan}
where they find large vibrations of the particles at lower frequency in the disordered structural regions.

This paper is organized as follows:
In Sec. II, we will explain the numerical methods on the oscillatory driven particle dynamics and the normal modes analysis. 
In Sec III, we will  show the results especially on the comparisons between the normal mode analysis and our new method.
In Sec IV, we will give the summary and discussion of the present study.

\section{Model and Simulations}

We consider a dense suspension of $N$ soft spherical particles at zero temperature in a two-dimensional square box of linear size $L$ with periodic boundary conditions on each side. 
The interaction between the particles is modelled by a short-ranged, purely repulsive, harmonic potential (similar to foams~\cite{durian}):
\begin{equation}
V(r_{ij}) = \frac{\epsilon}{2} \left( 1 - \frac{r_{ij}}{\sigma_{ij}} \right)^2 H(r_{ij} - \sigma_{ij}),
\end{equation}
where $r_{ij} = |\mathbf{r}_i - \mathbf{r}_j|$ and $\sigma_{ij} = (\sigma_i + \sigma_j)/2$. 
$\sigma_i$ and $\mathbf{r}_i$ are the diameter and position of particle $i$ respectively. 
$\epsilon > 0$ is the energy scale in the system and $H(x)$ is the heaviside function, defined such that $H(x) = 1$ if $x \ge 0$ and $H(x) = 0$ if $x < 0$.
The dynamics of each particle is described by the following equation of motion:
\begin{equation}
m\frac{d^2\mathbf{r}_i}{dt^2} + \xi\frac{d\mathbf{r}_i}{dt} = -\sum_{j=1,j\neq i}^{N}\frac{\partial V(r_{ij})}{\partial \mathbf{r}_i},
\label{eq:}
\end{equation}
where $N$ is the total number of particles in the system, $m$ is the mass of the particles and $\xi$ is the friction constant between the particles and the solvent.
We can define the natural frequency to be: $\omega_0=\sqrt{ \frac{\epsilon}{\sigma^2 m} }$ and
the damping coefficient to be: $\zeta = \frac{\xi\sigma}{2\sqrt{m\epsilon}}$
where $\sigma$ is the typical diameter of the particles. 
Physically if $\zeta>1$, the vibration of the particles will be overdamped, and if $\zeta<1$, the vibration is underdamped.
In this paper we shall fix $\zeta=0.01$ (underdamped regime) and $\omega_0=1$. 
Thus $\omega_0$ sets the timescale of our simulations.
The dynamical equation (\ref{eq:}) is integrated numerically using the velocity Verlet scheme~\cite{verlet}
(See Suppl. Info. for more detail).

The system is then forced out of equilibrium by oscillating each particle's diameter $\sigma_i(t)$ around its mean value $\sigma_i^0$~\cite{tjhung}:
\begin{equation}
\sigma_i(t) = \sigma_i^0 \left[ 1 + a\cos(\omega_dt + \psi_i) \right],
\label{eq:diameter}
\end{equation}
where $\omega_d$ is the driving frequency and $a$ is the amplitude of oscillation.
The average diameter $\{\sigma_i^0\}$ is taken from a bidisperse distribution of sizes $0.714\sigma$ and $\sigma$ with proportion $3:2$ to avoid crystallization.
We also introduce a fixed phase difference $\psi_i = 2\pi i/N$ for each particle $i$ so that the area fraction
$\phi = \sum_i^N \frac{\pi\sigma_i(t)^2}{4L^2} = \sum_i^N \frac{\pi(\sigma_i^0)^2}{4L^2}(1+a^2)$ is constant in time. 
Throughout this paper, we consider an area fraction close to jamming transition with  $\phi = 0.845>\phi_{\rm J} \sim 0.842$ \cite{lerner}
and a particle number $N = 1000$. 
Here, we use a small system size, because we calculate the Hessian matrix to compare the conventional normal mode analysis to our new method. 
In simulations, we also set the typical diameter of the particles as the unit of length such that $\sigma = 1$.

We start from a random initial configuration at time $t=0$.
We then introduce active oscillation in the particles' diameters according to Eq.~(\ref{eq:diameter}) with some fixed amplitude $a$ and 
driving frequency $\omega_d$.
If the amplitude of oscillation is large enough (above some critical value $a_c$),
the motion of the particles will be diffusive and the system will explore all the configurational space (ergodic) 
at steady state~\cite{tjhung,kawasaki16}.
On the other hand if the amplitude of oscillation is less than $a_c$ (the regime in which we are interested in),
the motion of all the particles will be periodic at steady state.
Thus in this vibrational state, the system is trapped in a local minimum of the total potential energy landscape.
The total potential energy is defined to be:
\begin{equation}
U(\mathbf{r}^N) = \frac{1}{2} \sum_{i=1}^{N} \sum_{j\neq i} V(r_{ij}),
\end{equation}
where $\mathbf{r}^N(t)=(\mathbf{r}_1(t),\mathbf{r}_2(t),....,\mathbf{r}_N(t))^T$ is a $2N$-components vector 
describing the 
configuration of the system at time $t$.
For a small oscillation amplitude $a<a_c$, 
the phase-trajectory of the system $\mathbf{r}^N(t)$ oscillates around some local minimum $\mathbf{r}^N_0$ at steady state.

At the local energy minimum $U(\mathbf{r}^N_0)$, the Hessian matrix is defined to be the second derivative of the potential energy with respect to the particles' positions (See Suppl. Info. for more detail):
\begin{equation}
\mathbf{H}_{ij} = \frac{\partial^2U}{\partial \mathbf{r}_i \partial \mathbf{r}_j} \Bigg|_{\mathbf{r}^N_0}.
\label{eq:hessian}
\end{equation}
The vibrational normal mode $\mathbf{e}_\omega=(\mathbf{e}_\omega^1,\mathbf{e}_\omega^2,....,\mathbf{e}_\omega^N)^T$ is associated 
to the eigen frequency $\omega$ according to the following equation:
\begin{equation}
\mathbf{H}\cdot\mathbf{e}_\omega = \omega^2 \mathbf{e}_\omega.
\end{equation}
Note that there is a set of $2N$ eigen frequencies $\{\omega\}$ and $2N$ corresponding eigen modes $\{\mathbf{e}_\omega\}$ for a given Hessian matrix.
The soft modes are the set of eigen modes $\{\mathbf{e}_\omega\}$ whose eigen frequencies $\{\omega\}$  are small compared to the natural frequency $\omega_0=1$.
These soft modes are usually characterized by long-range spatial correlation in the eigen vectors $\mathbf{e}_\omega^i$ of the particles.
Soft modes are typical of a disordered system close to jamming transition \cite{ohern, wyart05}.

In summary, from a given static configuration $\mathbf{r}^N_0$, we may calculate the Hessian matrix and subsequently a set of vibrational normal modes.
However this information is not sufficient to predict how the particles will move in real time.
Thus as one of the aims of this paper, we will make some comparisons between the vibrational normal modes and the real time particles' displacements.

\begin{figure}[ht]
\centering
\includegraphics[width=1.0 \columnwidth]{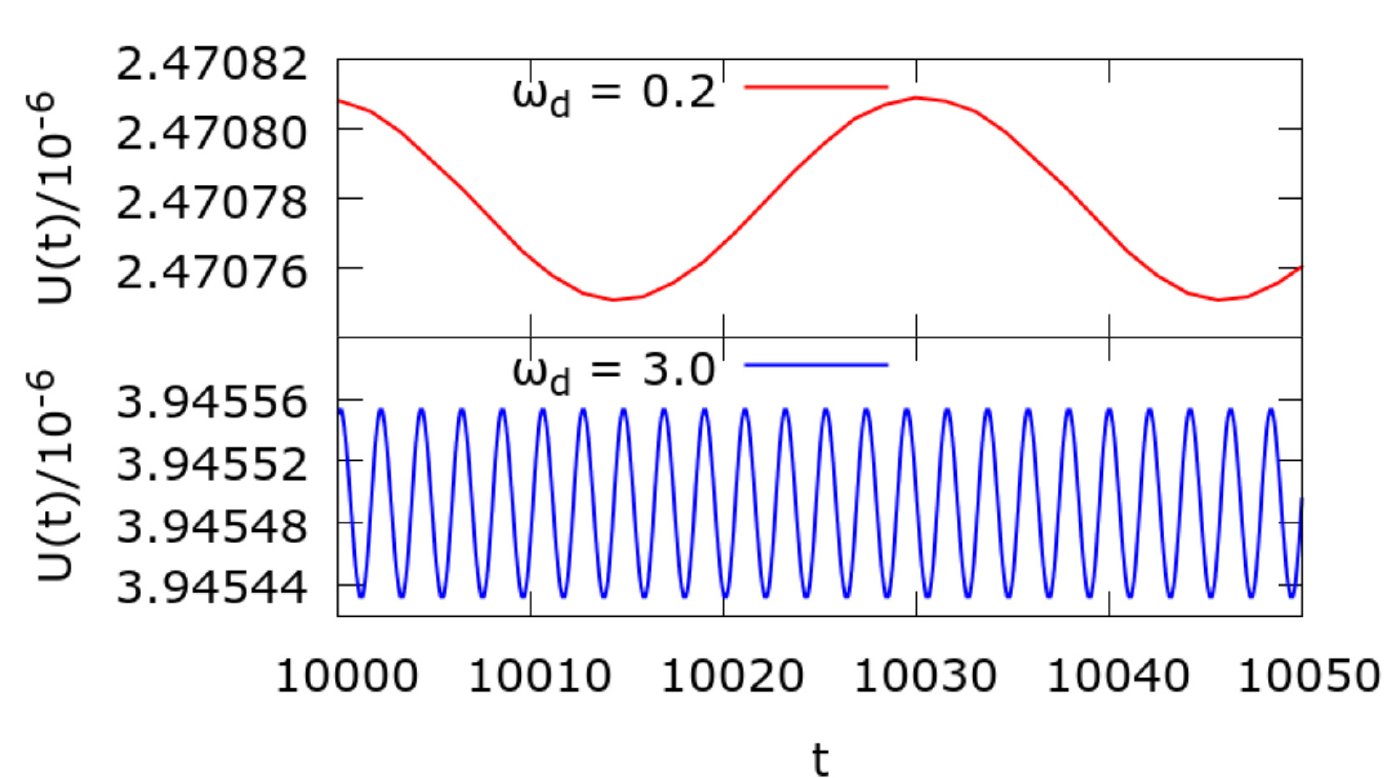}
\caption{
Above plots show the the total potential energy $U(t)$ as a function of time $t$ for two different driving frequencies: $\omega_d=0.2$ (top) and $\omega_d=3.0$ (bottom).
The amplitude of oscillation is kept fixed at $a=10^{-6}$ in both cases.
Steady state is already reached at $t=10000$ in the figure. 
At steady state, each particle vibrates periodically around its equilibrium position at frequency close to the driving frequency, thus,
the total potential energy $U(t)$ oscillates around some local minimum with frequency $\omega_d$.}
\label{fig:U}
\end{figure}

\section{Results}

Since we are only interested in the vibrational steady state,
we fix the oscillation amplitude to be much smaller than the critical amplitude $a_c$ ($a=10^{-6}$ in all our simulations).
We then vary the driving frequency $\omega_d$ from $0.1$ to $3.0$.
In this regime, the system will evolve into a vibrational steady state where all the particles move or vibrate periodically with a frequency close to the driving frequency $\omega_d$.
Therefore, the total potential energy $U(t)=U(\mathbf{r}^N(t))$ oscillates with frequency $\omega_d$ (see Fig.~\ref{fig:U}).
Note that in few cases, the system can also evolve into a vibrational steady state with a frequency double the driving frequency (period doubling)~\cite{reichhardt}.
We do not consider these cases in this paper since they occur much less frequently in the limit $a\rightarrow 0$.
Here, we also note that the configurations obtained from the vibrational steady states are not fully distorted with the active vibrations in any frequencies $\omega_d$ 
as the vibrational amplitude $a$ is extremely small.

\begin{figure*}[ht]
\centering
\includegraphics[width=1.0\textwidth]{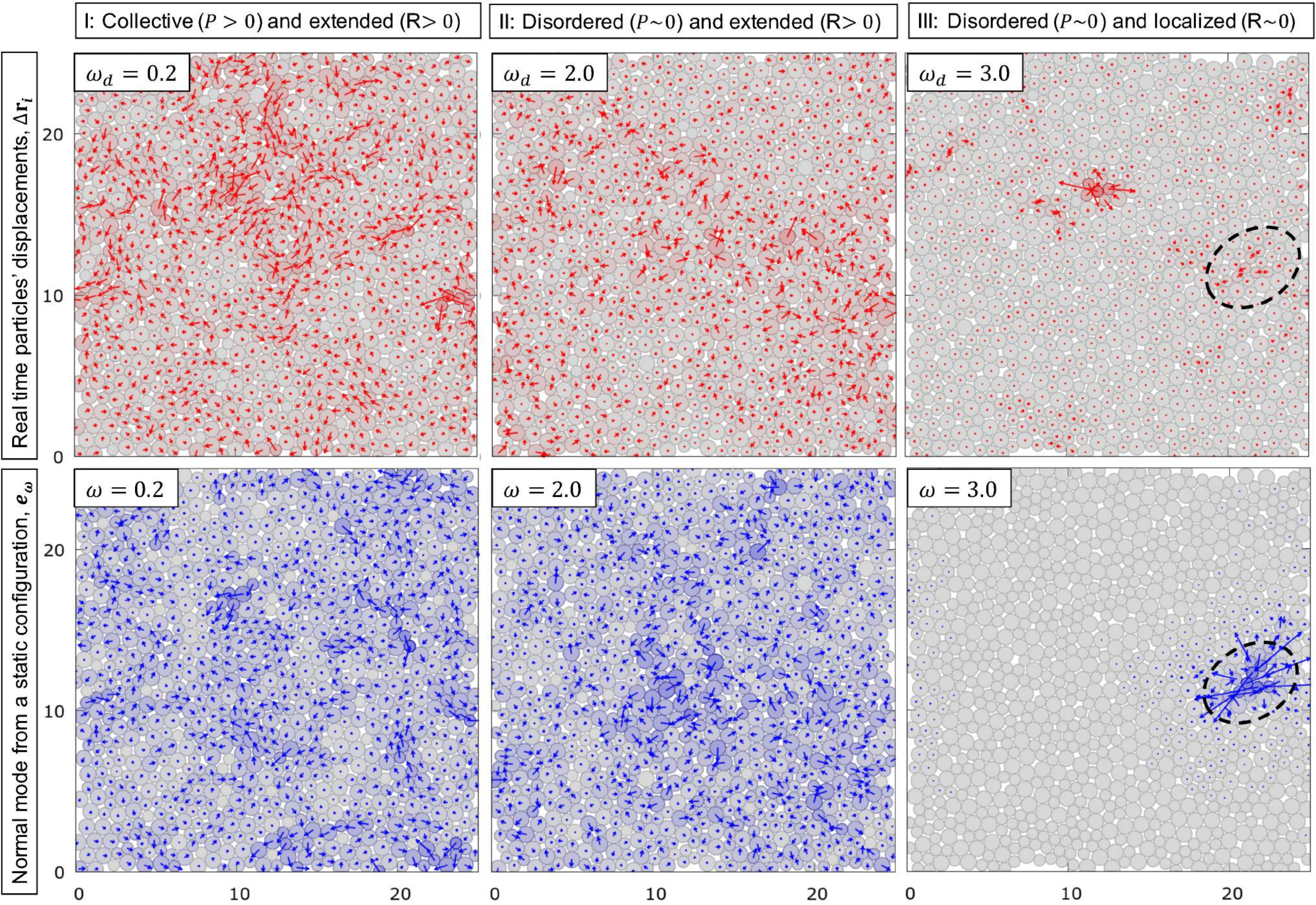}
\caption{
Top row: 
real time displacements of the particles: $\mathbf{r}_i(t+\Delta t) - \mathbf{r}_i(t)$ for increasing driving amplitudes $\omega_d$ from left to right.
(Here, $\Delta t = \pi/\omega_d$ or half period of oscillation and the oscillation amplitude is fixed at $a=10^{-6}$.)
For small driving frequency ($\omega_d=0.2$ or regime I), the displacements of the particles appear to be collective and extended over the whole space.
For intermediate driving frequency ($\omega_d=2.0$ or regime II), the  displacements of the particles appear to be more disordered but still cover the whole space.
Finally for high driving frequency ($\omega_d=3.0$ or regime III), the displacements of the particles become localized in space.
(The scale of the magnitude is around $10^{-5}$ in simulation units.)
Bottom row: 
vibrational normal mode (or eigen vector from the Hessian matrix) of a single static configuration for increasing eigen frequency $\omega$ from left to 
right.
It shows similar behaviour as the real time displacement field with three distinct frequency regimes.}
\label{fig:dr}
\end{figure*}

At steady state, we may look at the displacements of the particles between time $t$ and $t+\Delta t$:
$\Delta\mathbf{r}_i=\mathbf{r}_i(t+\Delta t) - \mathbf{r}_i(t)$,
where the initial time $t$ is larger than the time it takes for the system to reach a steady state and
the delay time $\Delta t$ is less than the period of oscillation $2\pi/\omega_d$.
(Note that the displacements of the particles are zero when $\Delta t = 2\pi/\omega_d$ since the motion of the particles is periodic.) 
Throughout this paper, we fix $t=10000$ and $\Delta t=\pi/\omega_d$ in simulation units (unless mentioned otherwise).
This delay time of equal to half the period of oscillation corresponds to the maximum displacements of the particles during one full oscillation cycle.
In Fig.~\ref{fig:dr} top row, we plot the real time displacement fields of the particles for increasing driving frequencies from left to right.
From these plots, we can distinguish three distinct frequency regimes based on two properties: collectivity and extensivity.
Here, we categorize the representative modes as performed in Ref.~\cite{hecke}, but with additional new definition of the collective measure. 
For small driving frequency or regime I ($\omega_d=0.2$ in Fig.~\ref{fig:dr}), 
the displacements of the particles appear to be collective and extended throughout the space.
However for intermediate driving frequency or regime II ($\omega_d=2.0$ in Fig.~\ref{fig:dr}), 
the displacements of the particles become disordered but still extended throughout the space.
Finally at high driving frequency or regime III ($\omega_d=3.0$ in Fig.~\ref{fig:dr}),
the displacements of the particles become localized in space and disordered.
See below for the quantitative definitions of these regimes by using the degrees of participation $R$ and polarization $P$ in the present study.
Here, we note that $P$ is a newly introduced degree of criterion, which has never been used in the previous studies.

\begin{figure*}[ht]
\centering
\includegraphics[width=1.0\textwidth]{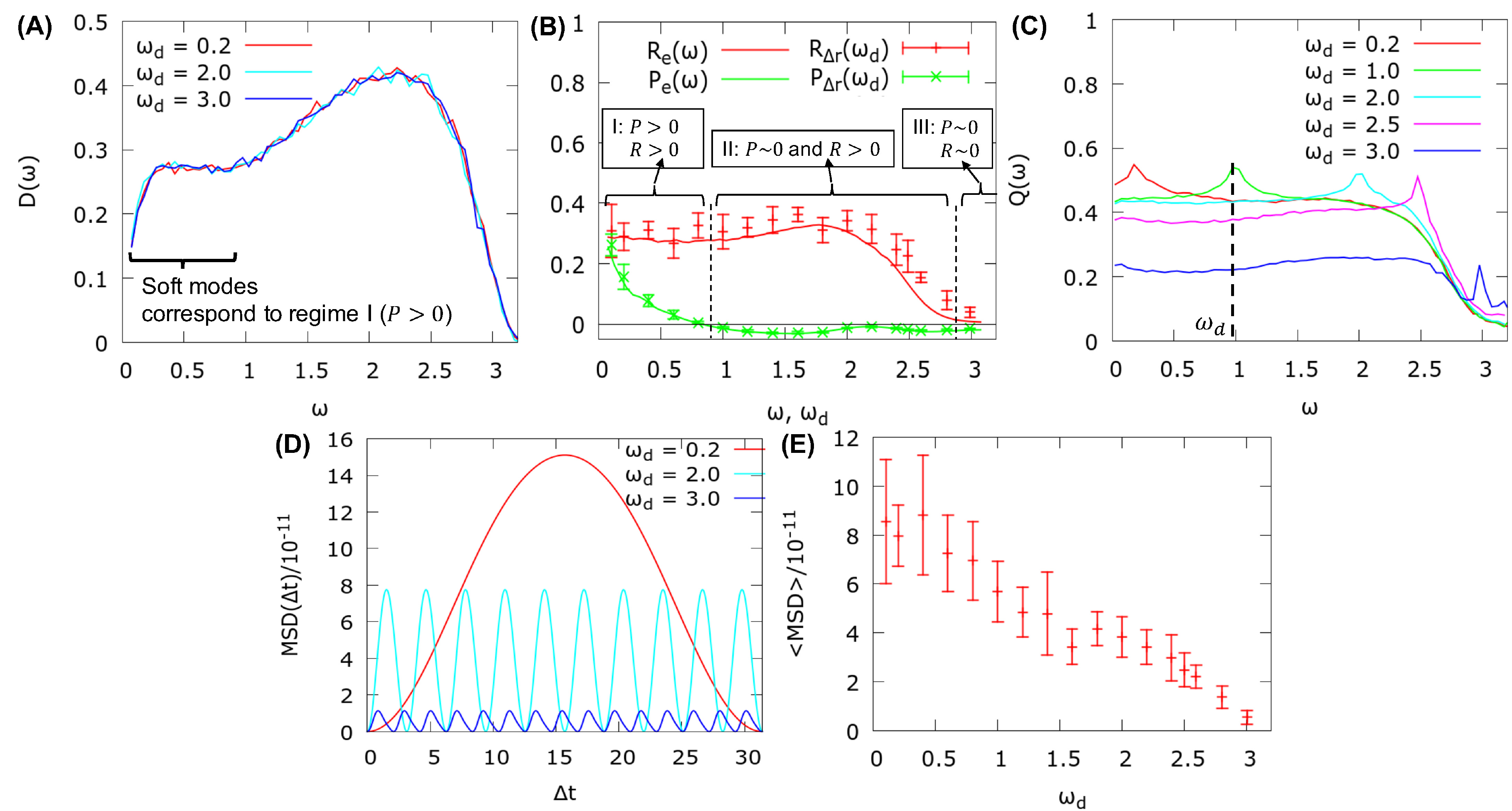}
\caption{
(A) shows the vibrational density of states $D(\omega)$ or probability distribution of the eigen frequencies $\omega$  for different driving frequencies $\omega_d$,
which is typical of a system close to jamming transition.
(B) The three frequency regimes from the displacement fields ($\Delta\mathbf{r}_i$) or eigen vectors ($\mathbf{e}_\omega)$ can be distinguished by plotting the participation ratio $R$ and polarization $P$
as a function of driving frequency ($\omega_d$) or eigen frequency ($\omega$) as introduced in Eqs. (9-11).
(C) shows the overlap function $Q(\omega)$, which is defined to be the dot product between the displacement field $\Delta\mathbf{r}_i$  
and the eigenmode $\mathbf{e}_\omega$ as introduced in Eq. (12).
(D) shows the mean squared displacement (MSD) as a function of delay time $\Delta t$ for different driving frequencies $\omega_d$
as introduced in Eq. (13) at the same amplitude of oscillation ($a=10^{-6}$).
The MSD is periodic as expected,
furthermore, the MSD for lower driving frequency is larger indicating the particles vibrate over larger distances compared to high frequency excitation.
(E) shows the average MSD over one period as introduced in Eq. (14), which increases as the driving frequency $\omega_d$ is decreased.}
\label{fig:dos}
\end{figure*}

To make comparison with the vibrational normal modes, we compute the Hessian matrix from a single static configuration at steady state.
More specifically, we drive the system with a fixed oscillation amplitude $a=10^{-6}$ and fixed driving frequency $\omega_d$ as before.
The value of the driving frequency $\omega_d$ is not important since they will all give the same distribution of normal modes as we shall see later.
We wait until the system reaches a vibrational steady state and then we take a snapshot of the system.
From this snapshot, we compute the Hessian matrix according to Eq.~(\ref{eq:hessian}), 
and after diagonalizing the Hessian matrix, 
we obtain a set of eigen modes (or normal modes) $\{\mathbf{e}_\omega\}$ and a set of corresponding eigen frequencies $\{\omega\}$.
Note that since the system is still evolving periodically, the configuration of the system at this instantaneous time is not strictly at the local energy minimum, 
and thus, we found a small fraction (around $1\%$) of imaginary eigen frequencies ($\omega^2<0$) which does not affect results that will be discussed below.
The distribution of the eigen frequencies or vibrational density of states (VDOS) $D(\omega)$ is plotted in Fig.~\ref{fig:dos}(A) for different driving frequencies $\omega_d$
(the data is averaged over $16$ independent simulations for a given $\omega_d$ and binned with bin size $=0.05$ on the $\omega$-axis).
$D(\omega)$ is normalized such that the area under the curve is equal to $1$.
As expected, VDOS does not depend on the driving frequency $\omega_d$.
It only depends on the natural frequency $\omega_0$ and packing fraction $\phi$ of the system, 
as long as we are sufficiently close to the local energy minimum (\emph{i.e.} $a$ is small enough compared to $a_c$).
Moreover from the plot, we can identify a region of soft modes which is typical of disordered system close to jamming density.

From the Hessian matrix above, we also plot the static vibrational normal modes $\mathbf{e}_\omega$ in the bottom row of Fig.~\ref{fig:dr} for increasing eigen frequencies $\omega$ from left to right.
Note that, the vibrational normal modes plotted in the bottom row of Fig.~\ref{fig:dr} are obtained from the static configurations directly above them.
For instance, the vibrational normal modes of $\omega=3.0$ (Fig.~\ref{fig:dr} bottom right) is obtained by diagonalising the Hessian matrix of the instantaneous particles' 
configuration in Fig.~\ref{fig:dr} top right.
Incidentally, we may also identify three distinct eigen frequency regimes depending on the collectivity and extensivity of the normal modes,
analogous to the real time displacement fields which we have observed in the top row of Fig.~\ref{fig:dr}.
Moreover, when we compare Fig.~\ref{fig:dr} bottom right to Fig.~\ref{fig:dr} top right, we may recognize a region of large overlap (dotted oval),
signifying some degrees of correlation between real time displacements and vibrational normal modes obtained from a static configuration.
Also note that when diagonalising the Hessian matrix, there exist several values of $\omega$'s which are very close to $\omega_d=3.0$.
Each of these eigen modes contributes to multiple localized regions of large displacements that we see in Fig.~\ref{fig:dr} top right (one of them is the region bounded by the dotted oval).

To quantify the collectivity of the displacement field $\Delta\mathbf{r}_i$ or the normal modes $\mathbf{e}_\omega$, 
we introduce a local polar order parameter $P^i$ in the neighbourhood of particle $i$.
More precisely, for the displacement field $\Delta\mathbf{r}_i$, the local polar order parameter $P^i_{\Delta\mathbf{r}}$ is defined to be:
\begin{equation}
P^i_{\Delta\mathbf{r}}(\omega_d) = \frac{1}{N_i} \sum_{j:|\mathbf{r}_j - \mathbf{r}_i|<\ell} \frac{\Delta\mathbf{r}_i \cdot \Delta\mathbf{r}_j}{|\Delta\mathbf{r}_i| |\Delta\mathbf{r}_j|}.
\end{equation}
where the subscript $j:|\mathbf{r}_j - \mathbf{r}_i|<\ell$ indicates summation over all particles $j$ whose distance from particle $i$ is less than $\ell$
and $N_i$ is the total number of such neighbours around $i$.
Here, we fix $\ell=3.0$ (in simulation units), although the result does not depend strongly of the value of $\ell$.
Physically $P\sim1$ if all the particles in the neighbourhood of particle $i$ move in the same direction and 
$P\sim 0$ if they all move in random direction (disordered or isotropic). 
Finally $P\sim -1$ corresponds to anti-ferromagnetic order which is not considered here.
Similarly we define the local polar order parameter for the normal modes $P^i_{\mathbf{e}}$ to be:
\begin{equation}
P^i_{\mathbf{e}}(\omega) = \frac{1}{N_i} \sum_{j:|\mathbf{r}_j - \mathbf{r}_i|<\ell} \frac{\mathbf{e}^i_\omega \cdot \mathbf{e}^j_\omega}{|\mathbf{e}^i_\omega| |\mathbf{e}^j_\omega|}.
\end{equation}
Finally, the average local polarization is then defined to be:
\begin{equation}
P_{\Delta\mathbf{r} / \mathbf{e}} = \frac{1}{N} \sum_{i=1}^N P^i_{\Delta\mathbf{r} / \mathbf{e}}.
\end{equation}
In Fig.~\ref{fig:dos}(B), we plot the average polarization of the displacement fields $P_{\Delta\mathbf{r}}$ (green points) and of the 
normal modes $P_\mathbf{e}$ (green line)
as a function of driving frequency $\omega_d$ and eigen frequency $\omega$ respectively.
(Again, the data is averaged over $16$ independent simulations and the $y$-errorbar is the standard deviation from the ensemble 
average. The bin size for the $\omega$-axis is $0.05$ as before.)
As can be seen from these plots, both $P_{\Delta\mathbf{r}}$ and $P_\mathbf{e}$ overlap each other, 
indicating high degree of correlation between the real time displacements and the static normal mode.
Furthermore, we also see $P\simeq0$ for all frequencies above $0.8$ in the plot, indicating disordered phase in the particles' 
displacements/eigen vectors in normal modes.
On the other hand, $P$ starts to increase (in the positive direction) as the frequency decreases below $0.8$ indicating increasing 
collective behaviour in the diplacement/eigen vectors. 
This coincides with the soft modes region as defined in jamming phenomena (see Fig.~\ref{fig:dos}(A)), 
and thus, we may also identify the soft modes to be the onset of collective order in the displacement or eigen vectors (regime I in our classification).

To quantify the degree of extensivity (or inversely, localization), we calculate the participation ratio $R$ as defined in~\cite{chen10,harrowell09}.
The participation ratio of the particles' displacements $R_{\Delta\mathbf{r}}$ is defined to be:
\begin{equation}
R_{\Delta\mathbf{r}}(\omega_d) = \frac{ \sum_{ij} \left| \Delta\mathbf{r}_i \right|^2 \left| \Delta\mathbf{r}_j \right|^2 }{N\sum_{i} \left| \Delta\mathbf{r}_i \right|^4}.
\end{equation}
Physically $R\sim1$ if the motion of the particles is extended over the whole system size, and
on the other hand, $R\sim1/N$ if the motion is localized.
Similarly, the participation ratio of the eigen vectors $R_\mathbf{e}$ is defined to be:
\begin{equation}
R_\mathbf{e}(\omega) = \frac{ \sum_{ij} \left| \mathbf{e}^i_\omega \right|^2 \left| \mathbf{e}^j_\omega \right|^2 }{N\sum_{i} \left| \mathbf{e}^i_\omega \right|^4}.
\end{equation}
In Fig.~\ref{fig:dos}(B), we also plot the participation ratio of the displacement fields $R_{\Delta\mathbf{r}}$ (red points) and of the normal modes $R_\mathbf{e}$ (red line)
as a function of driving frequency $\omega_d$ and eigen frequency $\omega$ respectively.
Again as before, both data almost overlap each other.
Finally, by studying both the participation ratio $R$ and the average polarization $P$, 
we can separate the three frequency regimes seen qualitatively in Fig.~\ref{fig:dr} more precisely:
regime I ($P>0$ and $R>0$),
regime II ($P\sim0$ and $R>0$), and 
regime III ($P\sim0$ and $R\sim0$).

We also compute the correlation between the real time particles' displacements (which are dynamical quantities)
and the normal modes of the Hessian matrix (which are static quantities).
When comparing Fig.~\ref{fig:dr} top right to Fig.~\ref{fig:dr} bottom right,
we have already seen a region of large overlap between the real time displacements and the eigen vectors (dotted oval in the figure).
Thus, we define the overlap function $Q(\omega)$ to be the dot product between the displacement field $\Delta\mathbf{r}_i=\mathbf{r}_i(t+\Delta t) - \mathbf{r}_i(t)$
and the eigen vector $\mathbf{e}_\omega$:
\begin{equation}
Q(\omega) = \frac{\left|\mathbf{e}_\omega\cdot\Delta\mathbf{r}^N(t)\right|}{\left|\mathbf{e}_\omega\right| \left|\Delta\mathbf{r}^N(t)\right|},
\end{equation}
where $\Delta\mathbf{r}^N=(\Delta\mathbf{r}_1,....,\Delta\mathbf{r}_N)^T$.
Note that the eigen vector $\mathbf{e}_\omega$ is obtained from the Hessian matrix of the particles' configuration at time $t$,
the initial time of the displacement vector $\Delta\mathbf{r}_i$.
Physically, the overlap function tells us how similar the displacement field is to the eigen vector $\mathbf{e}_\omega$.
$Q(\omega)$ varies from $1$ (maximum overlap) to $0$ (no correlation between $\Delta\mathbf{r}^N$ and $\mathbf{e}_\omega$).
We plot $Q(\omega)$ in Fig.~\ref{fig:dos}(C) for different driving frequencies $\omega_d$ from $0.1$ to $3.0$.
(Again, the data is repeated over an ensemble of $16$ independent simulations and then averaged.)
As can be seen from these plots, we observe a peak in the overlap function $Q(\omega)$ at exactly the driving frequency of the system $\omega=\omega_d$.
Thus by introducing an active oscillation in the system, we can excite the normal mode of the system which corresponds to the driving frequency of the active oscillation.

To quantify the relative magnitude of the particles' vibrations, we also compute the mean squared displacement (MSD)
as a function of delay time $\Delta t$:
\begin{equation}
\text{MSD}(\Delta t) = \frac{1}{N}\sum_{i=1}^N \left| \mathbf{r}_i(t+\Delta t) - \mathbf{r}_i(t) \right|^2,
\end{equation}
where the initial time $t$ is fixed. 
Fig.~\ref{fig:dos}(D) shows the MSD for increasing driving frequencies $\omega_d=0.2$, $2.0$ and $3.0$ for the same oscillation amplitude ($a=10^{-6}$).
As can be seen from the figure, the MSD is periodic with frequency $\omega_d$ as expected, however,
the amplitude of the MSD is larger for low frequency excitation ($\omega_d=0.2$) compared to the higher one ($\omega_d=3.0$).
In other words, the particles vibrate over larger distance when the driving frequency is lower.
This is consistent to the experimental results in~\cite{tan}, 
where they observe large low-frequency vibrational amplitude in the disordered structural regions.
This increasing distance over which the particles vibrate can be explained by increasing participation ratio and collective order as the frequency is lowered.

To analyze the dependence on the driving frequency more fully, 
we measure the average MSD over one period of oscillation:
\begin{equation}
\left< \text{MSD} \right> = \frac{\omega_d}{2\pi} \int_0^{2\pi/\omega_d}d\Delta t \, \text{MSD}(\Delta t).
\end{equation}
(Note that this quantity is also averaged over an ensemble of $16$ different simulations.)
We plot the averaged MSD as a function of driving frequency $\omega_d$ for a fixed driving amplitude $a=10^{-6}$ in Fig.~\ref{fig:dos}(E).
Indeed, we find that the averaged MSD does increase as we decrease the driving frequency.
(Note that the $y$-errorbar in Fig.~\ref{fig:dos}(E) is the standard deviation of the $\left< \text{MSD} \right>$ over the ensemble.)

\section{Discussions and perspectives}
In this study, we have constructed a new method to extract the specific vibrational modes as demonstrated in the normal modes analysis,
by using oscillatory driven particles in a two dimensional dense disordered system.
For the activation of the particle motions, we give a small amplitude of active oscillation in particle sizes with an identical driving frequency $\omega_d$.
First we have checked that the particles demonstrate reversible motion by means of the total potential energy 
for any input frequency $\omega_d$, as shown in Fig.~\ref{fig:U}.
Next we have shown the displacements of the driven particles for a half cycle ($\Delta t=\pi/\omega_d$) at several $\omega_d$ values in the top row of Fig.~\ref{fig:dr}. 
Then we have revealed that the distributions of displacement vectors remarkably depend on the values of $\omega_d$.   
For comparison, we have measured the eigen modes by means of the conventional normal mode analysis in the bottom row of Fig.~\ref{fig:dr}.
In particular, we observe overlapping regions between the real time displacements and the eigen vectors for the same frequency value, even though different techniques are applied. 
From the VDOS shown in Fig.~\ref{fig:dos} (A),
we have also checked that the normal mode analysis is not influenced by the external oscillations, because the oscillation amplitude is extremely small ($a=10^{-6}$). 
In Fig.~\ref{fig:dos} (B) we have presented  the degrees of spatial polarization and participation ratio for both methods,
presented respectively as $P_{\Delta\mathbf{r}/\mathbf{e}}(\omega_d/\omega)$ and  $R_{\Delta\mathbf{r}/\mathbf{e}}(\omega_d/\omega)$. 
It has been found that they behave respectively very similar, comparing the results obtained from both methods.
Furthermore we have measured the direct correlations between the real time displacement vectors of the particles driven with $\omega_d$
and the eigen vectors of the arbitrary modes $\omega$ obtained from the normal mode analysis by means of the overlap function $Q(\omega)$.
As shown in Fig.~\ref{fig:dos} (C), we have indeed found clear correlation where $\omega = \omega_d$. 
This means that our new method works well as performed in the normal mode analysis. 

From these analysis, we have revealed that both of the real time displacements for the oscillatory driven particles and 
the eigen vectors of normal mode analysis are composed of at least three different regimes such as 
regime I: the correlated extended regime ($\omega\sim 0.2$) due to $P_{\Delta\mathbf{r/e}}>0$ with $R_{\Delta\mathbf{r/e }}>0$, 
regime II: the disordered extended regime ($\omega\sim 2.0$) due to $P_{\Delta\mathbf{r/e} }\sim 0$ with $R_{\Delta\mathbf{r/e}}>0$, 
and regime III: the disordered localized regime ($\omega\sim 3.0$) due to $P_{\Delta\mathbf{r/e}}\sim 0$ with $R_{\Delta\mathbf{r/e}}\sim 0$
in the athermal amorphous solid close to $\phi_{\rm J}$.
Finally we have found that the typical amplitude of real time vibrations driven with low frequency oscillation 
is much larger than those with high frequency  by measuring the one cycle mean squared displacement $\langle \rm MSD \rangle $ as shown in Figs.~\ref{fig:dos} (D) and (E). 
These results are consistent with features of the eigen modes measured by the normal mode analysis~\cite{shintani08, tan}.

In this method, we are able to fully characterize the vibrational properties of amorphous materials 
(such as participation ratio $R(\omega)$ and polarization $P(\omega)$)
without the need of diagonalizing a large Hessian matrix.
Nevertheless, this method appears to be unable to directly measure the VDOS (or $D(\omega)$).
But, the VDOS has been obtained from the Fourier transform of the velocity auto correlation with respect to time 
without computing the hessian matrix~
\cite{rahman76,rahman81,ikedaJCP12,shintani08}.
Here, we note that the computational and memory costs for this velocity auto correlation together with the Fourier 
transform are of the order of $dN$.
Thus combined with our new method, we can obtain most of the basic vibrational quantities. 

It should also be noted that when we oscillate the particles' diameters,
the amplitude of oscillation should be kept small enough such that the response of the particles remains linear.
In our case, we always check that the total potential energy $U(t)$ is approximately sinusoidal with frequency equal
to the driving frequency for each simulation run. 
Previously we also defined $a_c$ to be the critical amplitude above which the motion of the particles become irreversible.
Below $a_c$, the motion of the particles is reversible, but not necessarily linear.
Nonlinearity in our system come in the form of period-doubling, in which,
the particles vibrate with a frequency $=\frac{\omega_d}{2},\frac{\omega_d}{4},\hdots$ ~\cite{reichhardt},
but the motion of the particles remains reversible.
In~\cite{tjhung}, finite size scaling shows that $a_c$ approaches to a finite value as $N\rightarrow\infty$.
Thus we can always find a reversible phase irrespective of the system size used. 
However in~\cite{schreck,van-deen}, it has also been shown that the window of linear regime with respect to $a$ 
in the reversible phase scales as $\sqrt{\Delta\phi}/N$, where $\Delta\phi= \phi-\phi_{\rm J}$.
This means that the linear regime vanishes at jamming transition point. And above jamming, the linear regime becomes 
narrower as $N\rightarrow\infty$.
Nevertheless, above jamming, we can always find a small enough value of $a$ in which the response is still linear for 
large but finite $N$.
In this paper we fix $a=10^{-6}$, but, we also perform several test simulations with $a=10^{-9}$ and we still find 
reliable results.This means we can easily extend our simulation to large system size $N\sim10^{6}$ and still find linear 
response with $a=10^{-9}$. Additionally, our method may also provide a useful tool to investigate the non-linear 
behaviour at large system sizes, close to jamming transition.

In future studies, by using our model, we will examine a much larger system size, 
not only for athermal jammed systems but also for thermal glasses. 
For instance in low temperature glasses, the scaled VDOS $D(\omega)/\omega^{d-1}$ 
shows a peak in the low frequency regime,
which is called the Boson peak. 
This has been observed in experimental systems such as oxide~\cite{ruffle06}, metallic~\cite{ruocco06}, and organic 
materials~\cite{ruffle08} as well as numerical simulations of model glasses~\cite{horbach01,shintani08}.
As a related matter, the low frequency vibrational modes in thermal glasses appear to be quasi-localized and collective
based on theoretical arguments~\cite{elliot},
as well as numerical simulations~\cite{schober91,schober93,schober96,harrowell08,harrowell09} and experiments~\cite{chen10}. 
According to the experiments of colloidal particles \cite{tan}, 
such characteristic behaviours of low frequency modes may originate from large vibrations due to structural defects, analogous to crystals.  
It is indicated that such low frequency modes could play a significant role in the appearance of Boson peak \cite{shintani08}. 
At even lower frequency, Debye modes also appear to be restored~\cite{elliott99}. 
To study such crossover behaviour, a much higher resolution of the frequency space will be needed, which requires 
much larger system sizes. 
However, conventional normal mode analysis requires a huge computational cost and thus a more efficient 
computational method will be desired.
To perform such studies, we propose a model where thermal noise is applied 
in addition to the active oscillations in the particles' diameters.
In particular, by using our analysis of both participation ratio $R(\omega)$ and average polarization $P(\omega)$,
we may identify multiple frequency regimes in low temperature glasses, 
which may coincide with the appearance of Boson peak at low frequency~\cite{chen10,harrowell09} 
as well as Debye modes at extremely low frequency~\cite{elliott99,xu}.

Our model can also be easily extended to other forms of two-body potentials such as Lennard-Jones, Gaussian core  {\it etc}.
In such case, we oscillate the interaction range and then we expect the same results.
It might also be interesting to consider other models of deformations which can excite a particular vibrational mode 
with a specific frequency.
For instance, we might apply an oscillatory shear on the system.
It is expected that the response of the system will correspond to the frequency of the oscillatory shear.
However in the case of oscillatory shear, the response of the system might not be as isotropic as compared to our mode of deformation.

Finally our model also represents a unique example of active matter where the system is driven with a characteristic driving frequency $\omega_d$.
Comparison of our actively oscillating particle system to the more mainstream self-propelled active particle system~\cite{henkes,berthier} might also be illuminating, 
since self-propelled particles do not possess such a characteristic vibrational frequency.

In conclusion, by means of our new method to excite the specific modes,
we have acquired results consistent with those of the normal mode analysis. 
Here the numerical costs for our new method is much lower than the normal mode analysis 
because we do not need to diagonalize a huge Hessian matrix. 
This makes us possible to perform vibrational analysis for much larger systems in near future, 
which will significantly contribute to the studies of the vibrational properties in any solids including soft materials.

\acknowledgments
We are grateful to Maxime Clusel who provided us with the initial ideas and for the illuminating discussions.
We thank Ludovic Berthier, Kunimasa Miyazaki, Akira Onuki, Atsushi Ikeda, Misaki Ozawa, and Kyohei Takae for variable discussions and comments.
TK gratefully acknowledges funding from JSPS Kakenhi (No. 15H06263 and 16H06018).
ET gratefully acknowledges funding from the European Research Council (ERC) 
under the European Union's (EU) Seventh Framework Programme (FP7/2007-2013)/ERC Grant Agreement No. 306845.

\end{document}